# The Width of the Gamma-ray Burst Luminosity Function

Andrew Ulmer[1] and Ralph A.M.J. Wijers[2]
Princeton University Observatory, Peyton Hall, Princeton, NJ 08544–1001, USA

## ABSTRACT

We examine the width of the gamma-ray burst (GRB) luminosity function through the distribution of GRB peak count rates, $C_{\rm peak}$, as detected by BATSE (BATSE Team 1993). In the context of galactic corona spatial distribution models, we attempt to place constraints on the characteristic width of the luminosity function by comparing the observed intensity distribution with those produced by a range of density and luminosity functions. We find that the intrinsic width of the luminosity function cannot be very well restricted. However, the distribution of intrinsic luminosities of *detected bursts* can be limited: we find that most observed bursts have luminosities that are in a range of one to two decades, but a significant population of undetected less luminous bursts cannot be excluded. These findings demonstrate that the assumption that GRB are standard candles is sufficient but not necessary to explain the observed intensity distribution. We show that the main reason for the relatively poor constraints is the fact that the bright-end part of the GRB flux distribution is not yet sampled by BATSE, and better sampling in the future may lead to significantly stronger constraints on the width of the luminosity function.

*Subject headings:* gamma-rays: bursts — methods: statistical

---

[1]E-mail: andrew@astro.princeton.edu

[2]E-mail: rw@astro.princeton.edu

## 1. Introduction

The problem of determining the luminosity function of GRB is a difficult one because there is no direct information about the distance to any individual burster. Consequently, the luminosity function can only be examined as an inverse problem involving the detected intensities and inferred spatial distribution. Assuming that there is no distance dependence in the luminosity function, we have

$$N(C_{\text{peak}} > C) = \int_{L_1}^{L_2} dL \int_0^{D(L)} dR \, 4\pi R^2 \rho(R) f(L) \qquad (1)$$

where $N(C_{\text{peak}} > C)$ is the number of bursts detected with peak count rate greater than C, $f(L)$ is the luminosity function, $\rho(R)$ is the spatial density function (with $R$ the distance from us), and $D(L) = \sqrt{(L/4\pi C)}$. Because only the intensity distribution, $N(C_{\text{peak}} > C)$ is observed, and the spatial distribution of the bursters is unknown, there is a natural ambiguity in the inverse problem.

However, much is already known about the form of the spatial density function. The observations of the *Burst and Transient Source Experiment* (BATSE) on the *Compton Gamma Ray Observatory* (CGRO) show an isotropic distribution on the sky in conjunction with a paucity of weak bursts relative to extrapolations based on $N(C_{\text{peak}} > C)$ of bright bursts (Meegan et al. 1992). In the region of space closer to us, the density is thought to be uniform as shown by, for instance, *Pioneer Venus Orbiter* (PVO) observations which have $N(> C) \propto C^{-1.5}$, as expected from a uniform density in Euclidean space independent of $f(L)$ (e.g. Fenimore et al. 1993). These observations have been shown to favor, among those density distributions which are spherically symmetric, a galactic corona or cosmological origin of the bursts (e.g. Mao & Paczyński 1992a,b). The Oort cloud has also been discussed (e.g. Clarke, Blaes, & Tremaine 1994). Within these distributions, it is possible to investigate the luminosity function. For example, Fenimore et al. (1993), in a cosmological scenario, determine the distance, and therefore intrinsic luminosity, of the bursts assuming a standard candle luminosity function. From these results, it follows that a standard candle is sufficient when fitting the peak flux distribution in either galactic or extragalactic models. The purpose of the present paper is to investigate whether it is also necessary. Previous work suggests that the effect of wide luminosity functions on the observed intensity distributions is small (Hakkila et al. 1993, Mao & Paczyński 1992b, and Hakkila et al. 1994). In contrast, it has recently been argued that intrinsic luminosity functions with width greater than about a factor of 10 are disallowed by a novel moment analysis of the intensity distribution (Horak, Emslie, & Meegan 1994). As we discuss below, we find similar constraints on the observed, though not intrinsic, luminosity function. A slight disparity in the degree of the constraints on the observed luminosity function can be understood as the result of differing samples and error analysis.

## 2. Method

There are many different ways of characterizing the number of events detected above a threshold count rate; the most popular method is $N(C_{\rm peak}/C_{\rm min} > X)$. Other distributions are closely related to this one, such as $V/V_{\rm max} = (C_{\rm peak}/C_{\rm min})^{-3/2}$. These methods involve the loss of some information in order to compensate for the low flux threshold effects. Furthermore, the $C_{\rm peak}/C_{\rm min}$ variable is not, in most cases, a perfect reflection of the intensity distribution (e.g. Petrosian 1993, Rutledge & Lewin 1993). We use data from the BATSE public catalogue (BATSE Team 1993). We adopt a simple method to assemble an $N(C_{\rm peak} > C)$ distribution that involves only a small loss of information while removing the threshold effects (a better, more elaborate method is given by Lynden-Bell 1971). We study only bursts with $C_{\rm peak}/C_{\rm min} > 1$ on the longest trigger timescale, 1024 ms. To account for threshold effects, a limiting minimum count rate, $C_\star$, is chosen; we remove from the subset all bursts with $C_{\rm peak} < C_\star$ or $C_{\rm min} > C_\star$ which correspond to very weak bursts and bursts which occurred during periods when the weak bursts in the sample could not have been detected. $C_\star$ is then varied to find the largest possible complete sample. With this selection process, the number of bursts in the subset is reduced to 165, from 193, the number of 1024 ms triggered bursts available. From this complete sample, an $N(C_{\rm peak} > C)$ curve is constructed. We note that the use of the statistic $C_{\rm peak}$ introduces a small effective width to the luminosity function because of varying incidence angles on the detector, but this width is only about a factor of two and does not strongly affect this analysis of the luminosity function (Rutledge & Lewin 1993).

For the analysis in this paper, we use spatial density distributions of the form

$$N(R) = \frac{n_{\rm o}}{(1 + (R/R_{\rm c})^\alpha)}, \qquad (2)$$

where $R$ is distance and $R_{\rm c}$ is the core radius. These distributions are commonly used to represent the distribution of matter in an extended galactic corona and are qualitatively similar to more complex cosmological models. Moreover, it has been shown that in the case of standard candles, galactic corona and cosmological models cannot be distinguished on the basis of the $N(C_{\rm peak} > C)$ distribution alone (Lubin & Wijers 1993).

With an observed $N(C_{\rm peak} > C)$ distribution and a parameterized family of density functions, we can investigate the luminosity function. Here, we consider a truncated power law distribution,

$$f(L) = CL^{-\beta} \qquad L_1 < L < L_2, \tag{3}$$

which has three parameters plus C as a normalizing constant. We examined specific subclasses of this luminosity function such as $L_2 = \infty$, $\beta = 0$, and $\beta = 1$; however, we found that the combination of these luminosity functions with the spatial density distribution function (Eq. 2) limited the outcome of our simulations *a priori*, because they constrained the fits to a region of parameter space with effectively narrow luminosity functions. The general power law functions alleviate this problem because they allow for extremely large widths around $\beta = 1.5$ while providing relatively good fits to the data. The best fit parameters for the luminosity function and spatial density functions are determined by maximum likelihood estimation so as to minimize the loss of information inherent to other techniques which require binning of the data. The likelihood function is

$$\prod_{i=1}^{n} -\frac{\mathrm{d}N(C_{\mathrm{peak}} > C)}{\mathrm{d}C}\bigg|_{C_i}, \tag{4}$$

where the model-determined function of number of detected bursts, $N(C_{\mathrm{peak}} > C_i)$, is given by equation 1 and $C_i$ is the peak count rate of the $i$th observed burst. This equation needs to be evaluated numerically except in a few cases. Equation 1 can be written in terms of incomplete beta functions and can be evaluated relatively efficiently for spatial density functions of the type (2). There are two parameters in the spatial density function and three in the luminosity function. One of these can be scaled out of the problem, because the characteristic distance, $R_c$, in the density distribution, and characteristic luminosities, e.g. $L_1$, only appear in the final expression through the combination $L/4\pi R_c^2$, a characteristic flux. In general, there are four free parameters that are determined using standard minimization techniques.

Once the parameters have been determined, we calculate the width of the *observed* luminosity function. We take the 90% width to be $L_{95\%}/L_{5\%}$. The luminosities $L_{95\%}$ and $L_{5\%}$ are such that the intervals $(L_1, L_{5\%})$ and $(L_{95\%}, L_2)$ each contain 5% of the total number of *detected* bursts so that:

$$0.05 = \frac{\int_{L_1}^{L_{5\%}} dL \int_0^{D_{\max}(L)} dR 4\pi R^2 \rho(R) f(L)}{\int_{L_1}^{L_2} dL \int_0^{D_{\max}(L)} dR 4\pi R^2 \rho(R) f(L)}, \tag{5}$$

$$0.05 = \frac{\int_{L_{95\%}}^{L_2} dL \int_0^{D_{\max}(L)} dR 4\pi R^2 \rho(R) f(L)}{\int_{L_1}^{L_2} dL \int_0^{D_{\max}(L)} dR 4\pi R^2 \rho(R) f(L)} \tag{6}$$

where $D_{\max}(L) = \sqrt{(L/4\pi C_\star)}$

## 3. Results

We find that in most cases, the intrinsic widths (i.e. $L_2/L_1$) of the luminosity functions cannot be limited. What can be somewhat limited is the width of the observed luminosity function. That is, the bursts that are observed primarily come from a limited range of luminosity, while the intrinsic luminosity functions may extend to weak bursts that cannot be detected. Figure 1 shows what limits can be placed on the width for different spatial density functions which range from a slow, smooth turnover at $\alpha = 2$ to an abrupt truncation of the sources for $\alpha = 8$.

A qualitative discussion of the shape of the flux distribution for different $\alpha$ (spatial density index) and $\beta$ (luminosity index) values will be helpful in understanding these results. A more general discussion of the flux distribution can be found elsewhere, (e.g. Wasserman 1992). First, define the critical fluxes $F_{1,2} \equiv L_{1,2}/4\pi R_c^2$, i.e. the flux of a source whose luminosity equals either end of the luminosity function as seen at a distance $R_c$. Many luminosity functions are effectively standard candles, so only geometry ($\alpha$) sets the shape of the flux distribution. Clearly, this occurs if $F_1 \approx F_2$; however, it is also the case when $\beta < 1$ or $\beta > 2.5$ because the luminosity function is then dominated by the luminous or weak sources, respectively. In all these cases the slope value is 0 if $\alpha > 3$, and $(\alpha - 3)/2$ if $0 < \alpha < 3$. The only case in which the luminosity function manifests itself clearly in the flux distribution is when $F_1 \ll F_2$, and the detector threshold is below $F_2$, and $1 < \beta < 2.5$. In these cases, the flux distribution is composed of a slope 0 or $(\alpha - 3)/2$ for $F \ll F_1$, a $1 - \beta$ slope for $F_1 \ll F \ll F_2$, and a $-3/2$ slope for $F_2 \ll F$. The maximum in Figure 1 occurs around $\alpha = 3$, because there the flux distribution has only the two latter components since the first is 0. To match the $-0.8$ low-flux slope seen with BATSE, this scenario requires $\beta \sim 1.8$, so the expected luminosity functions are often quite wide.

In short, two possible regimes exist within this general model that will fit the BATSE plus PVO data equally well: one is a standard candle (or effective standard candle) model with $\alpha \sim 1.4$ and a break to the uniform core near the upper flux regime of the first-year BATSE data. In these models, the effective width of the luminosity function –both observed and intrinsic– is small, less than a decade. The second is one with a wide luminosity function, $\beta \sim 1.8$, $F_2$ at the break to the 3/2 PVO slope, and $F_1$ below the BATSE threshold. Since the value of $\alpha$ only manifests itself below $F_1$, it is unconstrained by the data in this case, and since $F_1$ can be arbitrarily far below the BATSE threshold, the luminosity function can be arbitrarily wide intrinsically. But the observed luminosity function in this case may still be finite, because shifting $F_1$ down once it is below the threshold does not affect the width of the luminosity function of observed bursts. $F_2$ will be finite if a bend is convincingly detected in the data.

However, we only have the BATSE first-year catalog at our disposal. The sample we use was previously analyzed for deviations from a pure power law, and it was found (Wijers and Lubin 1994) that these deviations are still quite small, i.e. the bend to the 3/2 PVO slope is not very significant in it. This means that for a significant fraction of the bootstrap samples, $F_2$ will be poorly constrained and consistent with infinity. Then, the width ($L_{95\%}/L_{5\%}$) found for the luminosity function will become the maximum possible for the fitted value of beta: $(0.95/0.05)^{1/(\beta-1)}$. For $\beta$ close to 1.8, this is very sensitive to the value of $\beta$. Since for the sample size used (165 bursts) the two-sigma error in $\beta$ is 0.13 (when only fitting a slope), one can expect a range of widths from 25–100 to emerge from the fits. This only accounts for sample-to-sample slope variations in those cases where the sample is consistent with a power law. This problem is lifted for data sets in which the bend is well sampled, because then $F_2$ becomes well-constrained and its value limits the observed width to about a factor 10, irrespective of $\beta$. And this width limit, which we stress again only applies to the luminosity distribution of *detected* bursts, is set in essence by the distance from the BATSE threshold to the break in $\log N - \log C$.

The implication of the above for our fit results is that the typical width found for the observed luminosity function is substantial but not very large (20–40), but there is a heavy tail of very large widths (>100) in the distribution. Also, the distribution contains two types of values, those from near-standard candle fits with the BATSE slope being set by $\alpha$, and those with wider luminosity functions when the slope is set by $\beta$. (For each standard-candle model, one can find an alternative with a low-flux slope set by the luminosity function that is virtually identical to it over many decades on either side of the break in the slope.) The typically not too large width and the heavy tail towards very large width are apparent in our results (Fig. 2a).

Horack, Emslie, & Meegan (1994) find constraints on the width of the luminosity function using an entirely different technique. Their values are somewhat different from ours and employ different definitions for the width, a slightly different sample, and an error analysis in which Gaussian errors are assumed for the moments (whereas our results suggest that the widths have long non-Gaussian tails). We therefore believe that their numbers are consistent with our more conservative estimates. However, one should take care when interpreting their result, because they appear to claim that their constraint applies to the true or intrinsic luminosity function rather than the observed one, to which our results apply. However, their method is automatically limited to constrain only the luminosity function of detected bursts, even though they do not explicitly distinguish between the intrinsic and detected luminosity function in their paper.

For the future, it is natural to ask how the accumulation of more burst data will affect the constraints that can be placed on the distributions. To answer this question, a Monte Carlo simulation is used to produce intensity distributions from some best fit models to the observed data. Figure 2b-d are similar to figure 2a, except that the data sets are simulated from a model spatial density function and standard candles that fits the data. As more bursts are "detected" in the simulation, the constraints on the width of the luminosity function are strengthened. After about three years of operation, BATSE should have enough events to limit the 90% width to about 100 and the 80% width to about 40.

## 4. Conclusion

We find that we cannot constrain the width of the intrinsic gamma-ray burst luminosity function. The data are consistent with both narrow and wide intrinsic luminosity functions. However, weak constraints can be placed on the maximum range of luminosities of those bursts which are detected. Typically, the 90% confidence limit to the 90% width is a few hundred. We emphasize again that consistency of the luminosity function with standard candles is not surprising – it was found by many before. The new finding is that *the luminosities of detected bursts can also span a wide, yet limited range.* Given that most other measurable properties of the observed bursts have values ranging over many decades, this may already be a significant constraint on models in that they will have to produce a wide range of characteristics within a smaller observed range of luminosities. Also, models with wide luminosity functions leave open the possibility of alternative models recently discussed in the literature, such as wide luminosity functions due to relativistic sources beamed at varying angles relative to our line of sight (Brainerd 1994, Yi & Mao 1994).

This work was supported in part by NASA Grant NAG 5-1901. RW is supported by a Compton Fellowship (grant GRO/PFP-91-26). We thank B. Paczyński for helpful discussions.

---

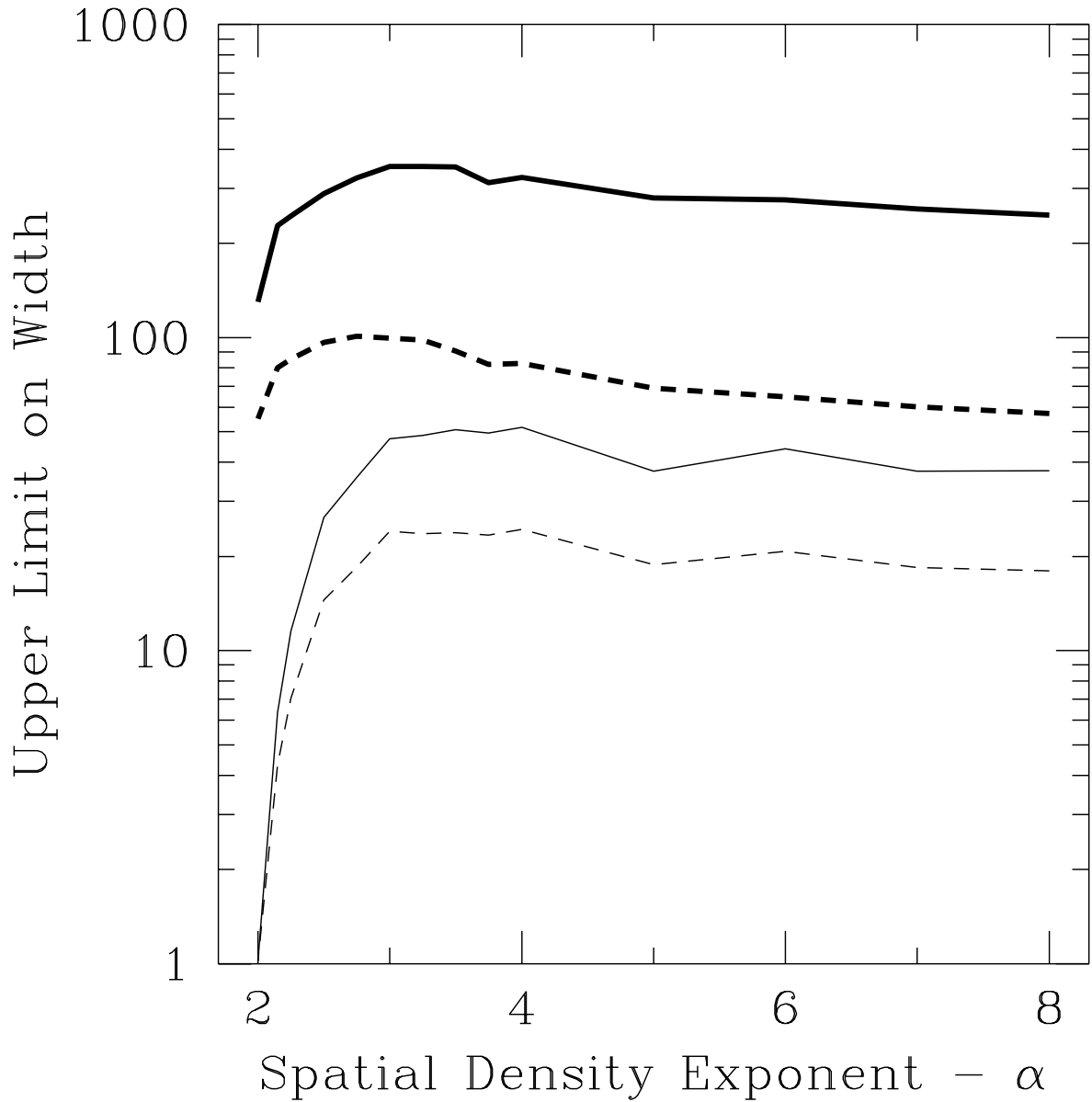

Fig. 1.— Constraints on the width of the observed luminosity function are shown over a wide range a spatial density functions. The top solid line corresponds to $L_{95\%}/L_{5\%}$ at 90% confidence. The next dashed line shows $L_{90\%}/L_{10\%}$ at 90% confidence. The lower solid line and dashed lines show median values of $L_{95\%}/L_{5\%}$ and $L_{90\%}/L_{10\%}$, respectively. The large difference between median and 90% confidence is an indication of the heavy tail in the distribution of widths.

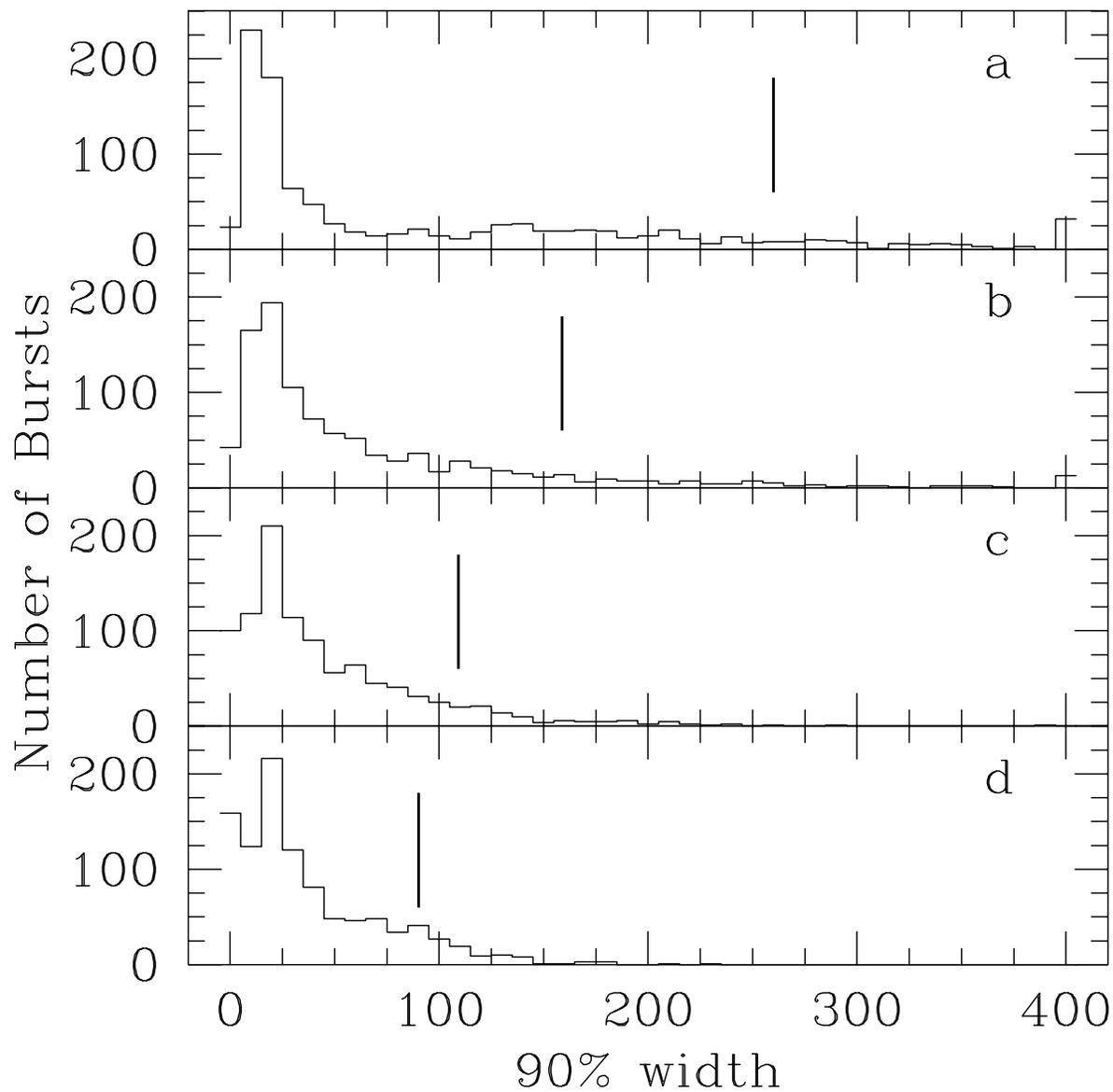

Fig. 2.— Figures 2a–d show distributions for the widths with $\alpha$, the spatial density index, as a free parameter. The 90% confidence upper limit to the $L_{95\%}/L_{5\%}$ are shown with vertical lines. Fig. 2a is composed from 1000 bootstrap resamplings of the data of 165 bursts each. Figures 2b–d illustrate possible future constraints. Shown are 1000 simulated observations of (165, 500, and 1000) standard candle bursts with a density distribution that well fits the data.